# THE PATTERN OF E-BOOK USE AMONGST UNDERGRADUATES IN MALAYSIA: A CASE OF TO KNOW IS TO USE

**Roesnita Ismail** [1] and **Zainab A.N.** [2]
[1] Faculty of Information Studies, Universiti Teknologi MARA, Malaysia
[2] MLIS Programme, Faculty of Computer Science and Information Technology
University of Malaya, Malaysia
e-mail: roesnita@salam..uitm.edu.my, zainab@um.edu.my

*ABSTRACT*

*This exploratory study focuses on identifying the usage pattern of e-books especially on how, when, where and why undergraduates at the Faculty of Computer Science and Information Technology (FCSIT), University of Malaya (UM), Kuala Lumpur use or do not use the e-books service provided by the University of Malaya library. A total of 206 (82%) useable questionnaires form the basis of analysis. The results indicate even though the students are heavy users of the Internet, rate themselves as skilled in Internet use and have positive attitude towards the e-book service, the level of e-book use is still low (39%). The students become aware of the e-book service mainly while visiting the University of Malaya Library Website, or are referred to it by their lecturers, friends or the librarians. About 70% rate positively on the e-book service. Those who are users of e-books find e-books easy to use and their usages are mainly for writing assignment or project work. Most respondents prefer to use e-versions of textbooks and reference sources. Generally, both users and non-users of e-books prefer to use the printed version of textbooks especially if the text is continuously used. There are significant difference between the frequency of e-book use and gender; between past usage of e-book and preference for electronic textbooks and reference books. The possible factors which may be related to e-book use are categorized into 4 groups and presented in a model, which comprises the ICT competencies of the students, their cognitive makeup, the degree of user access to the e-books and the functional or use factors.*

## INTRODUCTION

Some defined e-book as text that is available in an electronic format such as Word's doc, txt, HTML or XML (Hawkins, 2000; Ormes, 2002; McKnight and Dearnley, 2003; Vidana, 2003). Other definitions related to the conversion from print to electronic aspects of e-books, as printed text converted into digital form to be read on a computer screen (Saurie and Kaushik, 2001; Desmarais, 1994). Some definitions combined both the electronic text as well as the electronic reader device that is required for e-book to be read (Abrew, 2001; Lynch, 2001; Goh, 2003; Grant, 2002). Connaway (2003) defined e-book as a publication that characterized an electronic format, utilizing Internet technology to make it easy to access and use. Rao (2003) defined e-book as "text in digital form, or book converted into digital form, or digital reading material, or a book in a computer file format, or an electronic file of words and images displayed on a desktop, note book computer, or portable device, or formatted for display on dedicated e-book readers"

E-book databases are making its way into libraries through popular providers such as netLibrary, Books 24x7, Questia and Ebrary (Mullin, 2002; Connaway, 2003). Libraries on the other hand had begun to turn to providing e-book services simply because of decreasing budget, limited shelving space, increasing cost of new building and resources, the rising cost of repair or replacement of books, increasing demand from users for electronic resources, rising cost of inter-library loan service and the demand to support distance or distributed learning needs. E-books satisfied users' desires for immediacy and provided an easier and convenient access especially for remote users (McCarty, 2001; Snowhill, 2001). E-book subscriptions solved libraries' recurrent problems of lost or stolen or damaged books (Ardito, 2000; Connaway, 2003). Furthermore, the provision of e-books does not require unpacking, processing, shelving, and eliminate the extra time previously required to handle and process them before they can be used (Grant, 2002; Helfer, 2000). Features for handling e-books are also becoming user friendly, making it fairly easy for readers to browse, navigate, able to view graphics, videos or submit keywords and undertake full text searching within a book or a collection of books (Snowhill, 2001; Grant, 2002).

Through consortium subscription libraries could reduce their cost of purchasing per title, which would be formidable if a library embarked on lone subscription (Rohde, 2001; O'Leary, 2004). For instance, it would have cost 675 libraries participating in the TexShare database programme over US$167,741,000 to subscribe the database and e-book collection, which were purchased by the Texas State Library and Archives Commission for under US$8,000,000 (Ebrary, 2004). It would have cost each library US$89,266.000 if the database is subscribed individually. Ebrary charges academic libraries US$1.50 per full-time equivalent (FTE) student for the database of more than 13,000 titles. Ebrary also offers multi-year contracts and special pricing through consortia and regional networks (Ebrary, 2004).

Studies on whether, e-books are readily acceptable by users is still in its infancy. The University of California library (UCL) in February 2001 surveyed its four campuses and reported that most institutions were still in the trial stage with their e-book subscription. In that year UCL indicated that the acquisition of e-books had little or no impact on their purchase of printed titles. Librarians commented that they felt the role of e-books was not to replace printed text but serve as a duplicate copy (Snowhill, 2001). In Great Britain, Lonsdale and Amstrong (2001) indicated slow acceptance of nearly all digital textual resources other than journals. Helfer (2000) indicated that the slow acceptance may be due to users wanting to use e-books just as a reference resource, to look for the answers they want and sent the book back. Users of the SunShine library prefer to buy a copy of a book if they need it on an ongoing basis. Users of NetLibrary subscribed by the Associated Mennonite Biblical Seminary reported negative impression of the e-library among its users, who tended to just check out the database but had not actually read any books. The study also indicated those who actually use the NetLibrary gave positive ratings (Saner, 2002).

When using the e-book users prefer features such as glossary lookup, book marking, the ability to highlight and annotate in the e-book systems (Wearden, 1998; Simon, 2001). Simon's study of users at the Fordham College at Lincoln Center, Manhattan found that 84% of the respondents indicated being willing to contribute to the eBook retail cost. Gibbon (2001) reported a study carried out at the University of Rochester Libraries, New York State and found that only 29% (9 out of 31) respondents reported reading large portion of e-book titles and the rest either browsed or search for a single term across the collection. Gibbon attributed this situation to the discomfort of reading text on a computer screen for long periods of time. Gibbon's study however, revealed that when the e-book is a prescribed course reading, it tended to be used more frequently and this group of users also opined that it saved them money, it was convenient, it was accessible online even when the library was closed, and it did not require visiting the library. Similar responses were also given in Snowhill'study (2001). When comparing e-book use to users' ages, Anderson (2001), who surveyed 1500 US online Internet users found that those who frequent the Internet also rated e-book positively and they tended to be between the ages of 25 and 29. Some of the respondents indicated that e-books would be most useful if it is portable. A more recent study by Chu (2003) who surveyed 27 students in LIS schools in the United States in 2002 reported low use of e-book (9 out of 27). However, those who used e-book highlighted on the positive features such as their availability around the clock, search ability features and timely access to new titles.

Studies have related student's discipline to the rate of use of e-books. Dillon (2001) reported that students from the University of Texas at Austin mostly refer to e-books in the fields of economics, business and computer science followed by medicine and health. McCarty (2001) revealed that students at the University of Colorado Boulder Libraries used e-books for research and found it convenient, when searching for information. Healy (2002) interviewed 3200 faculty members, undergraduates as well as graduate students and observed that e-books were used for research, teaching and learning. A high percentage of users, however, still indicated preference for printed books and journals. Another study observed that publishing professionals preferred e-versions of reference materials, such as manuals, encyclopedias, maps and travel guides (Seybold, 2000). A study in the United Kingdom also found similar findings that consumers preferred e-books for reference purposes (Guthrie, 2002). Long (2003) also found that readers used e-books as reference sources or to look for a particular piece of information rather than reading from cover to cover.

The reasons for the slow acceptance of e-books vary but a constant reason given was not being comfortable with reading using personal computers, laptops and palm pilots (Helfer, 2000; Andersen, 2001). Other reasons included: finding it difficult to read on small screens, problems with browser, slow loading time, difficulties in navigating (Gibbon, 2001; Chu, 2003); and preferring to read printed text (Ray and Day, 1998; Holmquist, 1997; Gibbon, 2001). Summerfield and Mandel (1999) indicated that library users at the University of Columbia would use e-book in some depth when they are required to do so by courses they are following. The studies above indicated that the degree of acceptance of e-books is on the rise but the preference for printed text remained.

A survey of 14 Malaysian academic libraries web sites revealed only 6 provided e-book services for their users (International Islamic University Library; University of Malaya Library; Universiti Sains Malaysia; Universiti Utara Malaysia; Universiti Tun Abdul Razak; Open Universiti Malaysia). Most the libraries subscribed to NetLibrary (2), Ebrary.com (4), Books24x7.com (1); and xReferplus.com (1). Although there are a number of Malaysian academic libraries, providing e-books services, little information could be located to explain the current status and usage of e-books service. This study uses the case study approach in explore the use of e-book among Information Technology students from the Faculty of Computer Science and Information Technology, University of Malaya, which is subscribed by the University of Malaya Library (UML). The e-books have been included in the UML service since 2001 and can be accessed via the Library web site at http://www.umlib.um.edu.my. As at 2004, the Library has shared ownership of 2,457 e-book titles from Books24x7.com and more than 12,147 Ebrary titles on multi-disciplines.

**THE OBJECTIVES AND SAMPLE STUDY**

This is an exploratory study, which investigates the perception of e-books in term of its definition, the importance of e-books service at the UM Library and usage of e-books among IT students at the Faculty of Computer Science and Information

Technology (FCSIT), University of Malaya (UM). The objectives of this study are (a) To identify students' perception and understanding of electronic book in terms of what an 'e-book' is, and how they come to know about the e-books service at the University of Malaya Library; (b) To determine the usage of e-books among undergraduates with special attention on their impression of using e-books, the gateways used to access the e-books, the place where students access the e-books, the time spent to access the e-books, the subjects of e-books most used by students, the purpose(s) of using the e-books, the extent of e-book cited in the students' works, and the reasons for using e-books; (c) To identify non-users among the undergraduates and their reasons for not using; and (d) to identify students' preference of e-book to printed book.

The study population consisted of undergraduates studying in the second, third, fourth and fifth semesters for the academic year 2002/2003 at the Faculty of Computer Science and Information Technology (FCSIT), University of Malaya. They were selected for two reasons; firstly, it was assumed that the students have basic information literacy skills as they have attended the Information Skills course, which was offered during the second semester of their first year of study and secondly, being IT undergraduates, it was assumed that they were competent ICT users, would have little problems in handling a digital library environment and would more likely utilize electronic resources made available over the campus network.

A stratified sampling method was used to generate a random sample of undergraduates by programmes (Bachelor of Computer Science and the Bachelor of Information Technology) and semesters. Then, fifty undergraduates were selected based on an accidental sampling technique from the second, third, fourth and fifth semesters, bringing a total of 250 students. Each student was given a self-administered questionnaire as they entered their classrooms, computer laboratories, the faculty foyer, document room or canteen on a "first-come, first served" basis. A total of 206 questionnaires were returned, giving an 82% response rate. The data collected were analysed using Statistical Package for Social Science (SPSS) Version 11.0 for Windows. Suitable variables were tested for significance using the Pearson Chi-Square test ($x^2$). The students in this study comprised 94 (46%) male and 112 (54%) females.

**RESULTS AND DISCUSSION**

**(a) How is an E-book Perceived by Students?**
The majority of students either perceived an e-book as an electronic text or e-text. (55.8%) or as an e-text, which needed e-book devices and software to use (29.6%) (Table 1). These definition fits in with the definitions given by Desmarais (1994), Hawkins (2000), Saurie and Kaushik (2001), Ormes (2002), McKnight and

*Roesnita I & Zainab A.N.*

Dearnley (2003) and Vidana (2003) who focused on the contents (electronic text) of a book made available in an electronic format and Abrew (2001), Lynch (2001), Goh (2002), and Grant (2002), who equated e-books as a combined package of electronic text, e-book reading devices and e-book software. UM Library provided the e-book service published by Ebrary.com and Books24x7.com.

Only a small percentage of respondents indicated they did not know what an e-book was. The results show that the IT undergraduates of both genders were aware of the existence of e-books and gave acceptable definition to describe it.
.

Table 1: Perception of E-books (n=206).

| Definitions | Count | % |
|---|---|---|
| Electronic text (e-text) | 115 | 55.8 |
| E-text + e-book devices + e-book software | 61 | 29.6 |
| E-book software | 10 | 4.9 |
| E-book reader/devices | 8 | 3.9 |
| Don't know | 8 | 3.9 |
| No answer | 4 | 1.9 |
| **Total** | **206** | **100.0** |

**(b) How do Students become Aware of the E-book Service at UM Library?**
Figure 1 indicates that half of the students (50%) discovered information about the e-book service from the UM Library website, while about 17% were referred to it by their lecturers, fellow students (13%) and the librarians (13%). The brochures produced by the library to inform users of the service reached only a few of the undergraduates. There was no difference in the awareness pattern between male and female students.

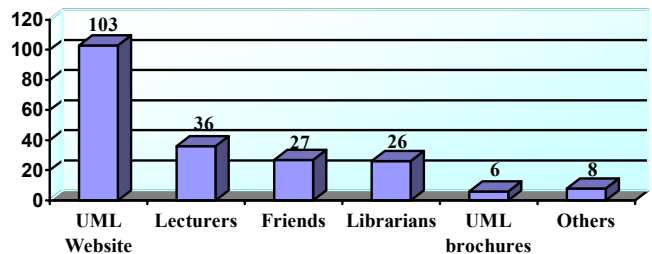

Figure 1: Means by Which Students Discover the UML E-book Service (n=206)

**(c) What Gateways was Used to Access the E-books?**
More than half of e-book users (65.4%) indicated that UM Library web site as the favourite gateway to access e-books, while the rest gained access through other libraries' web sites or other free websites or homepages (Table2). The high use of the UM Library web site is expected as UM Library homepage offered access to library collections from within the campus and remote users 24 hours a day and seven days a week.

Table 2: Gateways Used to Access E-books (n=81)

| Gateway Used to Access E-books | Count | % |
|---|---|---|
| Via the UM Library web site | 53 | 65.4% |
| Via the UM Library web site and other libraries' we site | 10 | 12.3% |
| Via other free Web/Homepages | 7 | 8.6% |
| Via the UM Library web site and other free web/ Homepages | 6 | 7.4% |
| Via other libraries' web site | 3 | .3.7% |
| All above | 1 | 1.2% |
| No answer | 1 | 1.2% |
| **Total** | **100** | **100%** |

**(d) From Which Location was the E-books Accessed?**
About 56% (45 out of 81) of those who used e-books access it from their Faculty computer laboratories as compared to their homes (25%, 20), UM Library (16%, 13) and other methods (3%, 3). These students therefore are taking advantage of the free and fast Internet access provided by the faculty to access the Internet and the UML websites. There was no significant difference in the location used to access e-books and gender.

**(e) What was the Student's Impression about Using the E-book?**
Out of the 206 respondents 81 (39%) indicated having used the e-book prior to the survey and out of this number only 17 (21%) indicated that they found it either not very easy or not at all easy. About 41 (51%) of users found e-book quite easy to use and the rest (27%, 22) found using it either easy or very easy. All 206 respondents were asked to give either negative, or positive ratings towards the e-book service and 79% (162) students rated "positive" or "very positive". Only about 2(1%) gave negative impression and 41(20%) were neutral. Those who gave negative reactions provided the following reasons for their impressions; they have used the e-book service prior to the survey and they still prefer to read printed text; they prefer to search for information via the Internet instead of using the e-book or they were not aware of the service or they were not interested or found it of no use to them. On the other hand, those students who have positive impression of e-book indicated that they found e-book services easy to access, easy to search for relevant information, convenient, economical, user friendly, time saving, and a good alternative service to library users. Seventy-two out of the 81 students who

*Roesnita I & Zainab A.N.*

have used the e-books indicated that they were willing to recommend the e-books to their colleagues and friends. The positive ratings applied to students of both genders. In an open-ended section of the questionnaire, a total of 54 statements were collected about the advantages of using the e-book and this is categorized and summarized in Table 3.

Table 3: Values Given to E-Books Services at the UML

| No. | Advantages | Total (n=54) | |
|---|---|---|---|
| 1. | Availability value | | |
| | Available online | 12 | 22.2% |
| | Easy access | 10 | 18.5% |
| | Accessible 24x7 | 7 | 12.9% |
| 2. | Enticing value | | |
| | Convenient | 4 | 7.4% |
| | User friendly | 3 | 5.6% |
| | Alternative service for library users | 2 | 1.9% |
| 3. | Handling value | | |
| | Easy to search | 3 | 5.6% |
| | Searchable | 3 | 5.6% |
| | Easy to cut and paste | 1 | 1.9% |
| 4. | Cost value | | |
| | Economical, free | 4 | 7.4% |
| | Time saving | 4 | 7.4% |
| | Save library space | 1 | 1.9% |

The results were similar with the findings of Ambikapathi (1999) on the information seeking behaviour of trainee teachers, who indicated accessibility as the most important criterion for choosing an information source while cost was found to be the least important factor. Moreover, time availability was seen as one of the primary issues surrounding the use of Internet and electronic resources rather than traditional library resources (Rice, 2003).

**(f) What was the Total Level of E-book Usage?**
The findings revealed that the majority of students (61%, 125 out of 206) have not used the e-books before the survey. This corresponded with Chu's (2003) findings who that reported that only a minority (33.3%) of his respondents have used e-books in the past. There was no significant relationship between the total use or non-use of e-books and gender even though in general more male students (44.7%) used e-books than female students (34.8%) (Table 4).

*The Pattern of E-book Use Amongst Undergraduates in Malaysia*

Table 4: E-books Usage by Gender (n=206)

| Used e-book in the past | | Gender | | Total |
|---|---|---|---|---|
| | | Male | Female | |
| No | Count | 52 | 73 | 125 |
| | % within Column | 55.3% | 65.2% | 60.7% |
| Yes | Count | 42 | 39 | 81 |
| | % within Column | 44.7% | 34.8% | 39.3% |
| Total | Count | 94 | 112 | 206 |
| | % within Column | 100.0% | 100.0% | 100.0% |

In terms of time spent, the results revealed that 72% (58) of the e-book users of both genders tended to spend between 5 hours to more than 7 hours accessing the Internet. However, this pattern is similarly exhibited by the non-user, where 80%(100) also spent between 5 and more than 7 hours a week on the Internet (Table 5). As a result, the chi-square test indicated no significant difference between the hours spent on accessing the Internet per week and the total use or non-use of e-books.

Table 5: E-book Use by Hours Spent on Online Accessing the Internet (n=205).

| Used e-books in the past | | Hours a week spend online accessing Internet | | | | | Total |
|---|---|---|---|---|---|---|---|
| | | < 1 hour | 1-2 hours | 3-4 hours | 5-6 hours | >7 hours | |
| No | Count | 1 | 8 | 16 | 37 | 63 | 125 |
| | % within Row | 0.8% | 6.4% | 12.8% | 29.6% | 50.4% | 100.0% |
| Yes | Count | 2 | 11 | 9 | 21 | 37 | 80 |
| | % within Row | 2.5% | 13.8% | 11.3% | 26.3% | 46.3% | 100.0% |
| Total | Count | 3 | 19 | 25 | 58 | 100 | 205 |
| | % within Row | 1.5% | 9.3% | 12.2% | 28.3% | 48.8% | 100.0% |

**(f) What was the Frequency of E-book Use?**
The students have not fully utilized the e-book service as the majority used the e-books service "occasionally" rather than weekly or monthly (Table 6). This may be due to the inability to access the e-books on daily basis as the computer laboratories, which form the location where most students accessed the e-book were often occupied for teaching purposes. Also, e-book use could be increased if librarians and lecturers refer students to e-books on their reading list and reference sources. The results also indicate that the frequency of e-books use among the users was significantly related to gender *(p* =0.016), where males are more frequent users of e-books compared to female students. This result concurs with the findings of Monopoli…et al. (2002) who found that more males used the e-journal services than their female respondents.

Table 6: Frequency of Use by Gender (n=79).

| Frequency over a period | | Gender | | Total |
|---|---|---|---|---|
| | | Male | Female | |
| Weekly | Count | 12 | 5 | 17 |
| | % within Column | 30.0% | 12.8% | 21.5% |
| Monthly | Count | 11 | 5 | 16 |
| | % within Column | 27.5% | 12.8% | 20.3% |
| Occasionally | Count | 17 | 29 | 46 |
| | % within Column | 42.5% | 74.4% | 58.2% |
| Total | Count | 40 | 39 | 79 |
| | % within Column | 100.0% | 100.0% | 100.0% |

$x^2 = 8.251$, df2, $p < 0.05$

When total hours spent on assessing the e-books per week was cross-tabulated with gender, the results showed no significant difference. The majority of both male and female students spent between an hour 3to up to 4 hours a week assessing the e-book (70%, 57 out of 81).

Table 7: Hours Spent Online Accessing E-Books Per Week by Gender (n=81).

| Hours Spent Accessing E-books | | Gender | | Total |
|---|---|---|---|---|
| | | Male | Female | |
| Less than 1 hour | Count | 13 | 8 | 21 |
| | % within Column | 31.7% | 20.0% | 25.9% |
| 1-2 hours | Count | 16 | 19 | 35 |
| | % within Column | 30.9% | 47.5% | 43.2% |
| 3-4 hours | Count | 10 | 12 | 22 |
| | % within Column | 24.4% | 30.0% | 27.2% |
| 5-6 hours | Count | 1 | 1 | 2 |
| | % within Column | 2.4% | 2.5% | 2.5% |
| More than 7 hours | Count | 1 | - | 1 |
| | % within Column | 2.4% | | 1.2% |
| Total | Count | 41 | 40 | 81 |
| | % within Column | 100.0% | 100.0% | 100.0% |

**(g) What are the Subjects of E-books most Used by Students?**
As expected, the majority of e-book users (71.6%) indicate that Computer Science and Information Technology as their favourite subject for use compared to "General" subject (23.5%), Health and Medicine (1.2%) and Economics and Business (1.2%). This was similarly found by Dillon (2001) whose respondents have also chosen more books in the same subject areas. These fields (science and technology) were attracting readers of e-books as in these fields current information is crucial and could be readily circulated through an electronic environment (UCONNLibraries, 2002). There was no significant difference in the types of subject use of e-book between male and female students.

**(h) What is the Primary Purpose of Students Using the E-books?**
The results found that students used e-books mainly for writing assignments/research projects (54.3%), reference (30.9%), leisure reading (6.2%), and browsing (3.7%) (Table 8). This is in agreement with the findings reported by McCarty (2001) and Guthrie (2002. This usage pattern of e-books applies to both male and female students.

However, the percentage of e-books cited by the 81 e-book users was found to be low. As 46.8% cited less than 10% while 34.2% utilized 20% to 40% and only 8.9% cited more than 50% of the e-books in their writings. The reasons for this may be related to the findings by Pew Internet & American Research Project (2000) that 71% of college students used the Internet as their primary source for their major school projects or reports. The minimal use of e-books in the students' works may also be due to their lack of information skills on how to cite electronic sources used in their writings. Students are not familiar with copyright regulations governing digital materials, and they are uncertain of the extent they can make copies of digital resources (Bodomo…et al, 2003). There was no significant difference in e-resources citations and gender of students.

Table 8: Percentages of E-Book Data Students Included In Their Works (n=79).

| Primary purpose of using the e-books | | Percentage of e-book data cited in students' works | | | | Total |
|---|---|---|---|---|---|---|
| | | None | < 10% | 20-40% | > 50% | |
| For assignments/ research projects | Count | 1 | 18 | 19 | 6 | 44 |
| | % within Row | 2.3% | 40.9% | 43.2% | 13.6% | 100.0% |
| For reference | Count | 2 | 19 | 4 | - | 25 |
| | % within Row | 8.0% | 76.0% | 16.0% | | 100.0% |
| For leisure reading | Count | 3 | - | 1 | 1 | 5 |
| | % within Row | 60.0% | | 20.0% | 20.0% | 100.0% |
| For browsing (e.g scan table of content) | Count | 1 | - | 2 | - | 3 |
| | % within Row | 33.3% | | 66.7% | | 100.0% |
| Other purposes | Count | 1 | - | 1 | - | 2 |
| | % within Row | 50.0% | | 50.0% | | 100.0% |
| Total | Count | 8 | 37 | 27 | 7 | 79 |
| | % within Row | 10.1% | 46.8% | 34.2% | 8.9% | 100.0% |

Both male and female (82.7%) respondents preferred to read the full chapter(s) or page(s) of e-books on the computer screens than the print format (16.0%). This corresponds with Monopoli…et al.'s study (2002) who reported two third of the respondents (69.5%) in the University of Patras, Greece, favoured the electronic version to read a journal article.

**(i) What are the Reasons for Using or Not Using the E-books?**
A high percentage of students indicated that they used e-book because it was available online (64.2%), provided faster and easy access to new titles (45.7%) and did not require physical visit to the library (40.7%) (Table 9). Bodomo…et al's (2003) respondents gave similar answers and his respondents recognized that digital libraries were very convenient since they did not need to go to libraries and could still read and download books or journals from home. Similarly, Chu (2003) also reported that "available around the clock" and "searchable" were valued the most by students at a library and information science schools in the USA.

Table 9: Reasons for Using or Not Using E-books

| **(a) Reasons for using (n=81)** | **Count** | **%** |
|---|---|---|
| Available online | 52 | 64.2 |
| Faster and easy access to new titles | 37 | 45.7 |
| Not require physical visit to the library | 33 | 40.7 |
| Easy to search | 31 | 38.3 |
| Convenient | 31 | 38.3 |
| Have user-friendly features | 17 | 21.0 |
| Available around the clock | 13 | 16.0 |
| Other reasons | 3 | 3.7 |
| No answer | 3 | 3.7 |
| **(b) Reasons for not using (n=125)** | **Count** | **%** |
| Prefer paper books | 57 | 45.6 |
| Little knowledge on how to use or access e-books | 44 | 35.2 |
| Inconvenient | 35 | 28.0 |
| Does not has Internet connection | 31 | 24.8 |
| Difficult to browse and read | 28 | 22.4 |
| No interest | 22 | 17.6 |
| Need special software | 12 | 9.6 |
| Other reasons | 2 | 1.6 |
| No answer given | 2 | 1.6 |

*Note: Students are permitted to give more than one reason.*

Table 9 (b) shows that almost half (45.6%) of the non-users indicate preference for paper format as a barrier for them from using e-books service. Holmquist (1997) found that the main reason for his respondents' non-use of e-journals was their preference to read articles on paper, not on the computer screen. Other non-users have mentioned factors such as little knowledge on how to use or access e-books,

*The Pattern of E-book Use Amongst Undergraduates in Malaysia*

the print copy is convenient to use, the lack of Internet connection, difficulty in browsing and reading, having no interest, and perceiving the need for special software to be able to use e-book as being cumbersome. When the non-users were asked whether they would use the e-book in the future, only 30% (38) gave a definite "yes" while the majority (61%, 76) indicated "probably" or "not sure or "probably not"" (6%, 8; 2%, 2).

This preference for printed text seemed to continue in the 2000s as found by McKnight and Dearnley (2003) during an investigation on the use of e-books in a public library in the United Kingdom. It is expected that the use of e-books would increase when respondents becomes more familiar with the e-books service. Mercieca (2003) described the reluctance to use electronic textbooks was due to respondent's perceived difficulty in reading electronic text. Students would only consider using them through the library collection primarily if there were no alternative printed texts.

**(j) Which Format did the Student Prefer: E-book Versus the Printed Book?**
When the total 206 respondents were asked to indicate their preference, the majority of students (81.1%) (both e-book users and non-users) indicated preferring to read printed to electronic textbook (18.4%) (Table 10). The results revealed that past usage of e-books is significantly related to preference for reading textbooks in the electronic text format ($p= 0.023$). Those who never use e-book in the past were more likely to prefer the printed text and a higher percentage of those who prefer the electronic format were those who have used e-books before. However, the 81 respondents who indicated actually using the e-book before the survey preferred to read the e-book on the computer screen (67, 83%) compared to those preferring to use the print version (13, 16%). This is similar to Rogers's (2001) finding that more students in American colleges (62%) preferred digitized textbooks to standard paper volumes

Table 10: Preferred Format for Textbook and Past Usage of E-Books (n=205)

| Preferred format for reading a textbook | | Used e-book in the past | | Total |
|---|---|---|---|---|
| | | No | Yes | |
| Print format | Count | 108 | 59 | 167 |
| | % within Column | 86.4% | 73.8% | 81.5% |
| Electronic format | Count | 17 | 21 | 38 |
| | % within Column | 13.6% | 26.3% | 18.5% |
| Total | Count | 125 | 80 | 205 |
| | % within Column | 100.0% | 100.0% | 100.0% |

$x^2$ =5.169, df1 , $p <0.05$

*Roesnita I & Zainab A.N.*

The situation may be different in an Asian environment, where, for example Bodomo…et al (2003) reported the majority of students (77%) in Hong Kong universities preferred printed materials. This result implied that the degree of exposure is important in increasing use of a product and as e-books was a recent addition to Malaysian libraries, its presence have not been fully explored by student users.

The majority of respondents (57%) equally preferred to use e-reference book. More than half of the respondents (57.3%) prefer to use a reference book in an electronic format. This study revealed that the past usage of e-books is very significantly related to preference of format for reading a reference book ($p=$ 0.0005) (Table 11). E-book users are more likely to prefer electronic reference books than the non-e-book users. This is found by Long (2003) whose readers use e-books as reference sources to look for specific information rather than for reading cover to cover.

Table 11: Using E-reference Books and Past Usage of E-Book (n=205)

| Preferred format for reading a reference book | | Used e-book in the past | | Total |
|---|---|---|---|---|
| | | No | Yes | |
| Print format | Count | 67 | 20 | 87 |
| | % within Column | 53.6% | 25.0% | 42.4% |
| Electronic format | Count | 58 | 60 | 118 |
| | % within Column | 46.4% | 75.0% | 57.6% |
| Total | Count | 125 | 80 | 205 |
| | % within Column | 100.0% | 100.0% | 100.0% |

$x^2 =16.334$, df1, $p <0.01$

Moreover, the additional features of e-books such as ease of browsing with keyword and full text searching capabilities within a book or across a collection of books helped enhance the usability of reference books such as dictionaries, encyclopedias and manuals. This study indicated that there was a very significant relationship between the preferred format for using a reference book to the preferred choice for reading e-textbook ($x^2 =16.447$, df1, $p <0.01$).  .

This may be true, as students who prefer to read printed textbook tended to use reference book in the similar format. The results of this study also show that many users and non-users of e-books believed that e-books would never replace the traditional printed books (108, 53%), while 23% (47) believed that electronic books will eventually replace printed text and 24% 950) were uncertain. Seybold Seminars and Publications (2000) reported that the majority of publishing professionals did not believe the use of printed text would diminish. Some academic librarians also felt the role of e-books was not to replace printed materials but to serve as a duplicate copy (Snowhill, 2001). As described by Helfer (2000), the users use e-books just as a reference tool to get the answer they needed and they could or would buy a physical copy of it if they wanted or need the book on an ongoing basis. This may also be true for the students in the present study.

**(k) How can UM Library Improve its E-book Service?**
Students volunteered suggestions on how to increase the utilization of the e-book service at the UM Library, which included; staff and librarians should encourage library users to try using the e-books, the UM Library should widely provide specific instruction on how to use or access e-books via the UM Library web site, the library should advertise new e-book titles on the UM Library web site, include e-books' titles in the library catalogs or OPAC and provide specific workshops on 'How to use e-book services' (Table 12). Some of these methods were tried at the University of Rochester libraries, where e-book titles were included within the library's catalogue, which resulted in an increase in use of the collection (Gibbon, 2001).

Table 12: Suggestions on Improving the E-books Service at the UML (n=206)

| Suggestions | Count | % |
|---|---|---|
| Motivate library users to try e-books | 100 | 48.5 |
| Provide specific instruction on how to use/access e-books via the UM Library web site | 99 | 48.0 |
| Advertise/promote new e-book titles on the UM Library web site | 89 | 43.2 |
| Include e-books' titles in library catalog/OPAC | 83 | 40.2 |
| Provide specific workshops on 'How to use e-book services' | 54 | 26.2 |
| Others | 5 | 2.4 |
| No answer | 3 | 1.4 |

Note: Students are permitted to give more than one answers.

**CONCLUSION**

E-book databases are recent additions to academic libraries in Malaysia and as such, little is known about its use or acceptance among educational users. This is a case study, which attempts to provide some information on the use of e-books among undergraduates at an academic library in Malaysia. The study is exploratory and aims to find out about students' awareness about the e-book services, types of use made of the e-books, reasons for use and non-use, the degree of e-books referenced in projects or essays and whether gender contributed to the differences in use and preference pattern of e-books. A total of 206 students from the Faculty of Computer Science and Information Technology, University of Malaya form the sample for this case study.

The results of the study revealed that students' use of e-book is greatly influenced by a number of interacting variables and this is indicated in a model given in

*Roesnita I & Zainab A.N.*

Figure 2. Use and non-use of e-books are determined by several circumstances comprising users' technological competencies; users own cognitive makeup; the level of access to e-books and the types of function or use made of the e-books.

The study assumes that ICT literate students would more likely use e-books. Although the majority of respondents have at least 1 to more than 7 years of computer experience prior to entering the degree programme (72%), more than 60% indicated that they have not used the e-books before the survey. The chi-square test also revealed that the number of hours spend on assessing the Internet per week is not related to use of e-books. This situation may indicate that although respondents are heavy user of the Internet and rate themselves as skilled in computer use, they hardly visit the library website or use the e-book service provided. This may be the reason for their ignorance of the service. Perhaps the surest way of making them use the e-book is to encourage academic staff to include e-book titles in their reading list as this method have been indicated to help increase use (Summerfield and Mandel, 1999). Also, academic libraries need to be creative in finding other means besides their websites and brochures to sell the service to users.

Students' cognitive makeup refers to their knowledge, perception, impression and attitude towards e-books. The students in this study seem to understand the concept of an e-book as most of them gave acceptable definitions. Most of the students (50%) discover information about the service from the UML website and the rest got to know via their lecturers, friends, librarians and library brochures. The students also rated positive towards this service even though they have not used it. The results indicate knowing about e-book and having a positive attitude towards e-book is not a definite determinant of use. However, it was found that past experience of e-book use is significantly related to e-book use, especially among male students, as more male students who are past users of e-book are also current users. Also, those who spend 3 to more than 7 hours accessing the Internet are also those who had used e-books in the past. This indicate that when students know about the service they would attempt to use and familiarity in using the service would most likely lead to current and future use. The libraries therefore, need to initiate student to use the service as the first step towards probable future use as past experience (especially if it is a positive one) would likely lead to subsequent use.

Access factors refer to situations that reveal how, where and why users access e-books. "How" infer the gateways used to access the e-books, "where" refers to the location where the service could be used and "why" refers to the purpose of use. The results reveal that UM library website is the most favourable gateway used to access e-books and the location of access is usually the faculty laboratories, the students' own homes and UM library. The results indicate that most of the e-book users are remote library users. Hence, making remote students aware of the "availability" of the service and user training is the first step towards attracting either on-site or online non-users when they attempt access to the service through the library website. Chu (2003) found that "availability around the clock", "easy search features" and "timely access to new titles" are reasons given by users who used e-books. Current student users are reluctant to visit the library. However, this should not be a hindrance for use as the campus networked environment allows for the use of resources remotely. Non-users of the current studies indicated preferring printed text and reported having little knowledge about e-books.

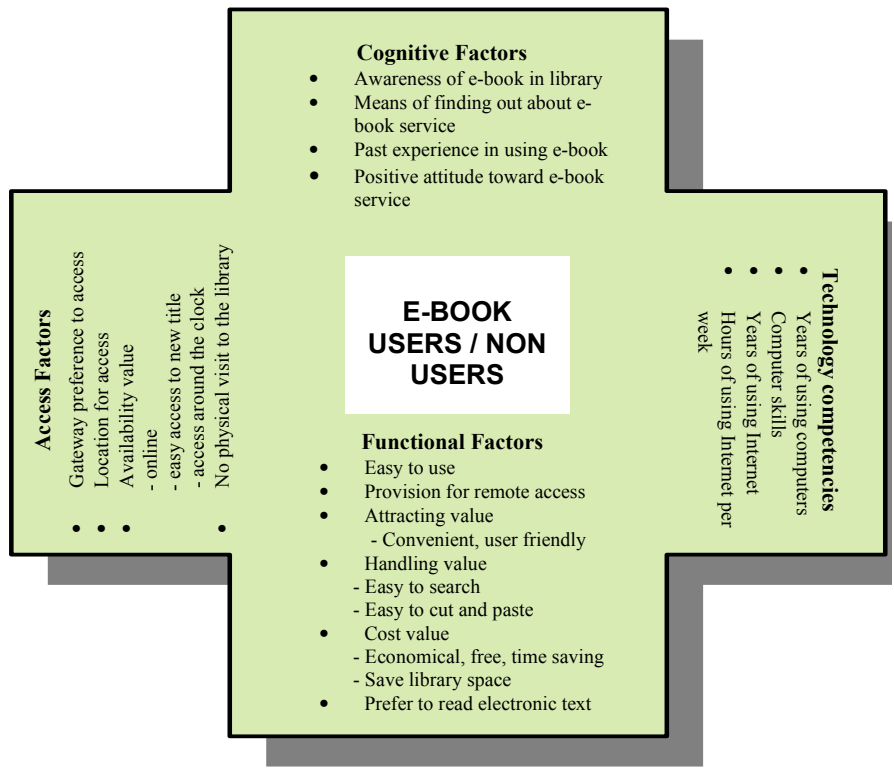

Figure 2: E-book Use Model

Online help guides and "Frequently asked questions (FAQ)" about e-books and other online services should be offered online and off line via the library web sites to allow easy access for all level of users. The library should also develop special links via the library web sites about library training sessions or workshops.

Simultaneously, libraries may also announce information about new library services and resources through the library or university newsletters.

Functional factors refer to situations that make it easy and cost effective for users to use e-books. The 81 users of e-books in this study indicate convenience, easy to access and user friendly as the criterion for using the e-books. The e-book users refer to e-book for their project work or assignments. They also prefer to use reference resources in the electronic form. Past use of e-book is found to be a significant factor in determining use of e-textbooks and preference for using e-book for reference purposes ($p = 0.05$). This results differ from the study of American college students by Rogers (2001) who indicated high preference for digitized textbooks (62%). The results indicate that the main reason for non-use is the preference for the printed text, especially if the text is continuously used throughout the academic year. In this instance the library can help by providing both print and electronic versions in the early stages of the services until users become more familiar in using the service. Both print and electronic text should be catalogued as separate entities and included in the library's catalogue. The user can choose to refer to whichever version that is available. This subsequently would help increase the usage of e-books.

In general, the results indicate that the level of e-book uses among the students was still low. The reasons for this are closely related to preference for the printed format and the lack of knowledge on its use. As such collection building based on user needs, selecting titles based on students required textbooks or reading lists, promoting activities to market e-book use needed to be carried out to create awareness among student users.

In this new environment, the librarian's role will change. Librarians need to be knowledgeable about e-books and tutor students on the options available when they are using the service. In order to become an effective instructor and intermediary for users, librarians should fully understand about the e-books service and always keep abreast of rapidly changing e-book technology. Continuing education about such service in the form of workshops has therefore become necessary. In order to encourage remote library users to use the electronic resources, library should incorporate individual titles in the library catalogue together with other electronic resources such as electronic journals, web pages, images and full text resources. By doing this, users are given the opportunity to search and retrieve relevant electronic-based resources through the web-based catalogues (Curtis, 2002).

The knowledge about the use of electronic services and resources becomes important in times of financial constraints and when evidence is needed to justify spending. Such knowledge becomes necessary when studying users' information needs and their information seeking behaviour, which are constantly changing. Student's use of e-book is therefore influenced by a number of interweaving variables, which are not static but continually related to each other. The framework provided could serve as an effective model for an understanding of the e-books' usage patterns among library users and address many of the library user-services related problems.